\documentclass[aps,pra, twocolumn,nofootinbib, showpacs]{revtex4-1} 
\usepackage[unicode=true,pdfusetitle, bookmarks=true,bookmarksnumbered=false,bookmarksopen=false, breaklinks=false,pdfborder={0 0 0},backref=false,colorlinks=false] {hyperref}
\hypersetup{ colorlinks,linkcolor=myurlcolor,citecolor=myurlcolor,urlcolor=myurlcolor}
\usepackage{braket,colortbl,amsthm,amsmath,amssymb,cleveref,txfonts}
\definecolor{myurlcolor}{rgb}{0,0,0.7}
\usepackage{graphics,graphicx}
\usepackage{color}
\usepackage{graphicx}
\usepackage[format = plain,labelfont = bf,up, textfont = normal , up, justification =raggedright, singlelinecheck =false]{caption}
\usepackage{subcaption}
\usepackage{tabularx}
\usepackage{amsmath}

\usepackage{graphicx}
\usepackage[utf8x]{inputenc}
\usepackage{color}
\usepackage{braket}
\usepackage{latexsym}
\usepackage{amssymb}
\usepackage{amsthm}
\usepackage{bm}
\usepackage{graphics,epstopdf}
\usepackage{color}
\usepackage{enumitem}
\usepackage{fmtcount}
\usepackage{booktabs}
\usepackage[normalem]{ulem}

\usepackage{epsfig}

\theoremstyle{plain}

\def\bea{\begin{eqnarray}}
\def\eea{\end{eqnarray}}
\def\ba{\begin{array}}
\def\ea{\end{array}}

\def\ket{\rangle}
\def\bra{\langle}
\def\beq{\begin{equation}}
\def\eeq{\end{equation}}

\begin{document}

%\title{Scaling of residual energy in adiabatic dynamics of disordered quantum Rabi model}%adiabatic or slow
%\title{Non-adiabaticity of a disordered dynamics close to quantum phase transition}
\title{Scaling of non-adiabaticity in disordered quench of quantum Rabi model close to  phase transition}

\author{Chirag Srivastava and Ujjwal Sen}

\affiliation{Harish-Chandra Research Institute, HBNI, Chhatnag Road, Jhunsi, Allahabad 211 019, India}

\begin{abstract}
Dynamics of a system exhibits non-adiabaticity even for slow quenches near critical points. We analyze the response to disorder in quenches on a non-adiabaticity quantifier for the quantum Rabi model, which possesses a phase transition between normal and superradiant phases. 
We consider a disordered version of a quench in the Rabi model, in which the  system residing in the ground state of an initial Hamiltonian of the normal phase is quenched to the final Hamiltonian corresponding to the critical point. The disorder is inserted either in the total time the quench or in the quench parameter itself. We solve the corresponding quantum dynamics numerically, and find that the non-adiabatic effects are unaffected by the presence of disorder in the total time of the quench. This result is then independently confirmed by the application of adiabatic perturbation theory and the Kibble-Zurek mechanism. For the disorder in the quench parameter, we report a monotonic increase in the adiabaticity with the strength of the disorder. 
Lastly, we consider a quench where the final Hamiltonian is chosen as the average over the disordered final Hamiltonians, and show that this quench is more adiabatic than the average of the quenches with the disorder in final Hamiltonian.

%In the realistic scenarios, it is almost impossible to perform the dynamics, ideally, hence it becomes important to investigate the non-adiabaticity of dynamics with the presence of disorder in the dynamics.
%In this article, we consider a disordered quench of the system which ideally initiates in the ground state of a normal phase Hamiltonian and end at the Hamiltonian corresponding to the critical point of QPT.

%For a sufficiently slow quench of the system from a normal phase Hamiltonian to the Hamiltonian corresponding to the QPT,
%the residual energy, a measure of non-adiabaticity of a quench, scales with the total time of quench maintaining a power-law, with the scaling exponent equal to $-\frac{1}{3}$.    

%We examine the linearly quenched dynamics of disordered quantum Rabi model. As the model exhibits a quantum phase transition, it is interesting to analyze its dynamics close to critical point where quantum phase transition occurs.  In specific, we study residual energy which measures the non-adiabaticity of the dynamics with the rate of quench. In case of a $slow$ quench rate, we observe that the prediction made by adaibatic perturbation theory and Kibble-Zurek mechanism does match the results obtained using exact solution. to the adiabatic dynamics of quantum Rabi model in the ordered case. 
\end{abstract}

%The existence of two quantum phases 
%prediction made -> scaling obtained

\maketitle
\section{Introduction}

Dynamics of quantum systems lead to many phenomena of fundamental importance in non-equilibrium physics \cite{meyer96,amico08,misbah10,silva'11,zudov12,eisert'15}. Recent advancements in the experiments have provided controlled implementation of dynamics in physical systems, like in cold atoms \cite{Schreiber15,Bordia16,Choi16,Bordia17a,Bordia17b} and ion-traps \cite{Smith16,Martinez16,Neyenhuis17,Zhang17}.
Inspecting the dynamics of quantum systems is crucial for technological advancements too, e.g., for modelling quantum simulators \cite{Lew07,Buluta09,Lew12,Nori14} and adiabatic quantum computations \cite{Cirac12,Lidar18}. Dynamics of physical systems possessing a quantum phase transition (QPT) has always been an interesting area of research \cite{Heyl13,Biroli11,Biroli13,Gambassi11,Silva15,Silva16,Heyl18,Halimeh18,Dutta18,Haldar20,Sehrawat20}. One striking feature of dynamics of a physical system near a  critical point 
%of a phase transition 
is that adiabaticity \cite{Born28} cannot be maintained due to a vanishing energy gap \cite{sachdev'11}. 

%Quantum phase transitions are usually considered to be the property of many body systems. Recently, it has been shown that a QPT may exist in a system even without going to the thermodynamic limit of the system size \cite{plenio'15}.

%For example, dynamics of systems near quantum phase transitions (QPTs) gives rise to non-adiabaticity due to a vanishing energy gap \cite{sachdev'11}.
%The observation of the criticality in QRM can also be seen using correlations and quantum entanglement of the system state.

%Inspecting the dynamics of quantum systems is crucial for technological advancements too, e.g., modelling quantum simulators and adiabatic quantum computations. Recent progress in the preparation of non-equilibrium states using cold atoms laid foundations for various experimental avenues as well.

  Quantum Rabi model (QRM), a simplified version of the Dicke model \cite{dicke'54}, with only one two-level atom interacting with a quantized cavity field, exhibits a QPT from a normal phase to a superradiant phase when the ratio of atomic transition frequency to the cavity field frequency goes to infinity \cite{bishop'96,bishop'01,levine'04,hines'04,feshke'12,nori'10,choi'10,ashhab'13,plenio'15}.  Thus in this case, the infinite ratio of frequencies plays the role of the thermodynamic limit in usual phase transitions.    Recently, it has  also been shown that the QPT in the QRM, as well as the dynamics near the QPT, can be experimentally probed in the setting of a single trapped ion \cite{plenio'17}.
In \cite{plenio'15}, a quench dynamics of the system described by a QRM in the normal phase %of the QRM 
is considered, and the non-adiabatic effect with respect to the rate of the quench is investigated.  For sufficiently slow quenches, a measure of degree of non-adiabaticity of a quench, the residual energy, follows a power-law scaling with the quench rate.  For quenches ending away from the critical point, the fall in residual energy is sharper than for the quench ending at the critical point. For slow enough quenches, the ideas of adiabatic perturbation theory (APT) \cite{grandi'10} and the Kibble-Zurek mechanism (KZM) \cite{kibble'76,zurek'85,zurek'96,zurek'97,dziar'10,zurek'14} can be applied to obtain the scaling of residual energy with respect to the rate of quench. 
The KZM was initially developed to describe formation of the universe, and later on applied to condensed matter systems.
It predicts formation of defects when the system is driven through a symmetry breaking phase transition. It has  been successfully applied to various models
possessing phase transitions
\cite{dicke'54, lipkin'65,botet'82,botet'83,polko'05,zurek'05,dziar'05,damski'05,damski'06,damski'07,damski'08,niko'16}.

%Note that recently it is shown that the QPT in QRM as well as its dynamics can be probed in settings of a single trapped ion.

%Quantum systems driven out of equilibrium, has been widely explored and still is an area of active research.  Dynamics of quantum systems lead to many phenomena of fundamental importance in the non-equilibrium physics.  Dynamics of systems near quantum phase transitions (QPTs) may give rise to phenomena like universality and non-adiabaticity. 

%Inspecting the dynamics of quantum systems is crucial for technological advancements too, e.g., modelling quantum simulators and adiabatic quantum computations. Recent progress in the preparation of non-equilibrium states using cold atoms laid foundations for various experimental avenues as well.

In real circumstances, the system parameters and the devices controlling the tuning of these parameters may not be ideal. In such scenarios, it is important from an experimentalist's point of view to know about the ``average'' behavior of relevant physical quantities. 
In this article, we study the quench dynamics in the QRM in presence of a disorder in the quench parameters. The aim is to analyze the (quenched) average non-adiabatic effects of the dynamics, affected by quenched disorder.
Ideally, the system residing in the ground state of the Hamiltonian in the normal phase  is linearly quenched to a final Hamiltonian corresponding to the critical point of the QPT. We analyze  disordered versions of the above quench in the following two cases. In the first case, the disorder is present in the total quench time, whereas in the second case, there is disorder in the quench parameter. 
%final Hamiltonian does not correspond to the critical point.

In both the cases, we study the average residual energy, where the average is taken over all disordered quenches.
For the first case, we find that the quenched average residual energy scales similarly as the residual energy for the ordered quench, with respect to the rate of the ordered quench. Therefore, the non-adiabaticity for quenches ending at the critical point, is unaffected by the disorder in the rate of quench. For sufficiently slow quenches, we also apply APT and KZM to obtain the same result, which indicates the applicability of KZM even to disordered quenches. For the second case of disordered quench, we report a monotonic increase in the adiabatic nature of the quench with the strength of the disorder. 
Furthermore, we also consider a quench where the final Hamiltonian is chosen as average over disordered final Hamiltonians. We show that this quench is ``more adiabatic'' than the quenches with the disordered final Hamiltonians.

 The rest of the paper is organized as follows. In sec. \ref{sec2}, we discuss the prerequisites, and is subdivided into three subsections. We start with briefly reviewing QPT in the QRM in subsec. \ref{subsec2a}. In subsec. \ref{subsec2b}, dynamics of the QRM and a measure of non-adiabaticity of the quench is discussed, and in subsec. \ref{subsec2c}, quenched averaging of physical quantities over the disorder is briefly explained. In sec. \ref{sec3}, we study a quench of the system from a given initial Hamiltonian to the critical point, but with a disorder in the rate of quench. For sufficiently slow quenches, we use APT and KZM to reproduce the scaling exponents in subsec. \ref{subsec3a}. In sec. \ref{sec4}, we study a disordered quench of the system from a given initial Hamiltonian to a disordered Hamiltonian which do not correspond to the critical point Hamiltonian. We present a conclusion in sec. \ref{sec5}.

%The degree of non-adiabaticity with the rate of a linear quench, ending away to critical point or ending at critical point, is also studied. For slow quenches ending away from a critical point 

%QRM nonadiba...KZM
%Quench dynamics crossing the QPTs are extensively studied for sudden as well as slow quenches.
%Slow quenches to critical points results in non-adiabatic evolution.

%Its interesting to study dynamics in a system near a phase transition. Specially, adiabatic evolutions fail at points close to phase transitions. There exist models in which Kibble-Zurek mechanism (KZM) is applied and has been successful in predicting the scalings. But there also exist models where its prediction fail. Quantum Rabi model (QRM) is shown to undergo a quantum phase transition in an infinite atomic to photonic frequency limit. Dynamics of QRM in this limit is studied before. For slow  linear quench, scaling of non adiabaticity function is observed. It is further shown that the prdictions made using adiabatic perturbation theory (APT) and  KZM matches with the exact evolution for slow quenches. In this letter we study the dynamics of disordered Rabi model. Interestingly, we observed that APT and KZM used in disordered case to study the scaling of non-adiabaticity function fails to verify the numerical exact solution.
\section{Prerequisites}
\label{sec2}

\subsection{Quantum Rabi model}
\label{subsec2a}
 We briefly discuss here about a quantum phase transition in quantum Rabi model. 
A two-level atom in the presence of quantized cavity field can be modelled by the quantum Rabi Hamiltonian, which has the form 
\begin{equation}
H_{Rabi}=\hslash\Big[\omega a^{\dagger}a  +  \frac{\Omega}{2}\sigma_z - \lambda (a^{\dagger}+a) \sigma_x \Big] ,
\label{eq.1}
\end{equation}
where $\hslash$ is the reduced Planck's constant, $\omega$ is the frequency of the cavity field, $\Omega$ is the atomic transition frequency, $\lambda$ is the coupling coefficient, $\sigma_{x,y,z}$ are the usual Pauli operators, and $a$ and $a^{\dagger}$ are respectively the annihilation and creation operators of cavity field. 
%We represent eigenstates of $\sigma_z$ by $|\!\!\uparrow\ket$ and $|\!\!\downarrow\ket$, such that $\sigma_z|\!\!\uparrow\ket=|\!\!\uparrow\ket$ and  $\sigma_z|\!\!\downarrow\ket=-|\!\!\downarrow\ket$. Fock basis elements, $|n\ket$, where $n$ belongs to the set of whole numbers, are eigenstates of number operator, $a^\dag a$.
 Exact solutions to quantum Rabi Hamiltonian are devoid of any closed form, despite it being solvable~\cite{braak}. It is shown in \cite{plenio'15} that an effective Hamiltonian describing low energy physics of $H_{Rabi}$ can be deduced for  $\frac{\Omega}{\omega}\gg 1.$ In fact, in the limit $\frac{\Omega}{\omega},\frac{\lambda}{\omega}\to\infty$, $H_{Rabi}$ undergoes a quantum phase transition, which can be shown using its effective low energy Hamiltonian \cite{plenio'15}.

The effective low energy Hamiltonian for $H_{Rabi}$ in the limit $\frac{\Omega}{\omega}\to\infty$ is given by
\begin{equation}
  \begin{aligned}
H_{np}&=\hslash\Big[\omega a^{\dagger}a  - \frac{\omega g^2}{4}(a^{\dagger}+a)^2 - \frac{\Omega}{2} \Big]; \text{  }\left|g\right| \leqslant 1, \\
H_{sp}&=\hslash\Big[\omega a^{\dagger}a  - \frac{\omega}{4g^4}(a^{\dagger}+a)^2 - \frac{\Omega}{4}(g^2+g^{-2})\Big]; \text{  }\left|g\right|>1,
  \end{aligned}
  \label{eq.2} 
\end{equation} 
%
%H_{sp}=\hslash\Big[\omega a^{\dagger}a  - \frac{\omega}{4g^4}(a^{\dagger}+a)^2 - \frac{\Omega}{4}(g^2+g^{-2})\Big] \text{ if }\left|g\right| >1
  where $g=\frac{2\lambda}{\sqrt{\Omega\omega}}$ is a dimensionless parameter and  quantum phase transitions occur at $|g|=1$. Subscripts $np$ in $H_{np}$ and $sp$ in $H_{sp}$ stands for normal  ($|g|\leqslant 1$) and superradiant  ($|g|>1$) phases, respectively.  
% In the mentioned limits, the energy spectrum of $H_{Rabi}$ gets divided into two sectors. The lower energy sector corresponds to the subspace of $|\!\!\downarrow\ket$ and hence, $H_{np}=\bra\downarrow\!\!|H_{Rabi}|\!\!\downarrow\ket$.  
In order to obtain the low energy effective Hamiltonians from $H_{Rabi}$ in the limit $\frac{\Omega}{\omega}\to\infty$, certain unitary transformations are applied such that the low energy spectrum of $H_{Rabi}$, corresponds to one of the spin subspaces. Then a projection of $H_{Rabi}$ on that spin subspace provides the effective Hamiltonians \cite{plenio'15}.

%The exact eigenenergies of the effective normal phase Hamiltonians are given by $\hslash \epsilon_{np}a^{\dagger}a+E_{G,np}$. Since, we are intersted in the dynamics of the model in the normal phase only, for details see \cite{plenio'15}. 
%Note that the spin operators, corresponding to atomic part of atomic-photonic system,  are absent in the hamiltonians in Eq. (\ref{eq.2}). This is because, while deriving Eq. (\ref{eq.2}) from Eq. (\ref{eq.1}) in the mentioned limits~\cite{plenio'15}, unitary transformation of $H_{Rabi}$ are demanded for which its energy spectrum gets divided in low and high energy sectors based on spin subspace, respectively. Then the low energy effective hamiltonians in Eq. (\ref{eq.2}) are obtained by projecting the transformed Rabi hamiltonian in low-energy spin subspace.  
%$U^{\dag}H_{Rabi}U$ 

%\subsection{Linear quenched dynamics in the normal phase of the QRM}

%Define dynamics, define non adiabaticity in terms of residual energy, mention previous results, Verification by  APT + KZM.

\subsection{Linearly quenched dynamics in normal phase}
\label{subsec2b}
In this subsection, we briefly present the quench dynamics of the system, prepared initially in its ground state, with a linearly quenched time-dependent  Hamiltonian in the normal phase \cite{plenio'15}. Let $g$ be the quench parameter and have the form
\begin{equation}
\label{lata}
g(t)=\frac{g_f}{\tau}t;\hspace{0.5cm}0\leq t\leq \tau.
\end{equation} 
%This implies that t
The dynamics starts at time $t=0$ with $g=0$, and ends at time $t=\tau$ with $g=g_f$.  
%As mentioned, at time $t=0$, the system is prepared in its instantaneous ground state.
Here, $g_f$ is the final value of the quench parameter, $g$, and $\tau$ is the total time of quench. Note that $g_f\leq 1$ for system to be in normal phase and $\tau$ controls the rate of the quench from $g=0$ to $g_f$. Larger the values of $\tau$, the slower is the rate of the quench. It is known generically that the evolution of any system, if it is always gapped, can be adiabatic if the quench rate is sufficiently slow. But for a dynamics very close to a critical point, adiabaticity breaks even for very slow quenches.
To capture the non-adiabaticity of such dynamics, let us consider a quantity called the residual energy, given by
\begin{equation}
E_r(\tau)=\bra\Psi(\tau)|H_{np}(g(\tau))|\Psi(\tau)\ket-E_G(\tau),
\end{equation}  
where $|\Psi(t)\ket$ is the wavefunction of the system at time $t$ and $E_G(t)$ is the ground state of instantaneous Hamiltonian $H_{np}(t)$. 

Solving the dynamics for Hamiltonian $H_{np}(g(t))$ (using equation of motion in the Heisenberg picture), reduces to solving the following differential equations \cite{plenio'15}
\begin{eqnarray}\label{heisen}
\frac{i}{\omega} \frac{du(t)}{dt}=\left(1-\frac{g^2(t)}{2}\right)u(t)-\frac{g^2(t)}{2}v(t), \nonumber \\
-\frac{i}{\omega} \frac{dv(t)}{dt}=\left(1-\frac{g^2(t)}{2}\right)v(t)-\frac{g^2(t)}{2}u(t),
\end{eqnarray}
with initial conditions $u(0)=1$ and $v(0)=0$, and the constraint $|u(t)|^2-|v(t)|^2=1$.
And the residual energy is then given in terms of $u(t)$ and $v(t)$ as
\begin{equation}
E_r(\tau)=\hslash\left(\omega|v(\tau)|^2-\frac{\omega^2 g_f}{4}|u(\tau)+v(\tau)|^2-\frac{\omega\left(\sqrt{1-g_f^2}-1\right)}{2}\right).
\end{equation}
%\begin{figure}[h]
%\includegraphics[width = 0.3\textwidth, angle=-90]{Ervstq.eps} 
%\caption{The residual energy $E_r$ is plotted against $\omega\tau$ for different final quench parameter $g_f$. For slow quenches ($\omega\tau \gg 1$), $E_r$ scales as $\tau^{-2}$ and $\tau^{-1/3}$, respectively, for the quenches ending sufficiently away from the critical point and for the quench ending at critical point. $E_r$ is given in the units of $\hslash \omega$, and $\omega \tau$ is dimensionless.  
%}
%\label{fig1}
%\end{figure}

It has been shown that for very $slow$ quenches ($\omega\tau \gg 1$), $E_r$ scales with total time of quench, $\tau$, maintaining a power law. For quenches ending far away from the critical point, i.e., $g_f \ll 1$,  $E_r \propto \tau^{-2}$, whereas for quenches ending near the critical point, i.e., $g_f\sim 1$, $E_r \propto \tau^{\frac{-1}{3}}$ \cite{plenio'15}.
%(see Fig. \ref{fig1}).  
The scaling exponent makes transition from -2 to $\frac{-1}{3}$ as the end point of quench moves towards the critical point, when the excitations in the system starts to be appreciable (i.e., adiabaticity breaks). 

\subsection{Quenched averaging}\label{subsec2c}
In real experiments, inaccuracies invariably creep in. 
Therefore, it becomes important to look out for a quantity of interest under such scenarios. In this paper, our aim is to study quantities affected by disorder in a quench dynamics of the system. The relevant quantities of interest are therefore disorder averaged quantities. However, the type of averaging depends on the strain of the disorder.
Consider $O$ to be an observable of interest for a given quench dynamics of the system. And let $\delta$ be a parameter, which brings in disorder for a given single run of the quench dynamics, and 
%is chosen from a 
the corresponding probability distribution is given by $p(\delta)$. Let the disorder-affected observable be given by $O_\delta$. Then, if the disorder in \(\delta\) is of the quenched type, viz. if the equilibration time of the disorder is much larger than the time required for all the relevant observations about the observable \(O\), then the quantity of interest for an experimentalist
%in such scenarios 
will be the quenched average of $O$, 
%over such disordered quench dynamics, 
which is given by
\begin{equation}
\overline{O}=
%\sum_{\text{all~}\delta\text{~values}}p(\delta)O_\delta.
\frac{1}{[\delta]}\sum_\delta O_\delta,
\end{equation}
where the sum runs over all realizations of the disorder, obeying the distribution \(p(\delta)\), for different runs of the experiment, and \([\delta]\) is the number of runs of the experiment. One of course needs to perform a convergence check using a higher number of runs of the experiment.

\section{Dynamics with disorder in total time of quench}
\label{sec3}
In this section, we consider a disorder in the rate of the quench, i.e., in the total time of quench. We
study the effect of disorder on the scaling of the disorder averaged residual energy, $\overline{E}_r$, with respect to the quench time, $\tau$, of the quench in the ordered case.
We consider a quench where the system starts in the normal phase ($\frac{\Omega}{\omega}\to\infty$ and $|g|<1$) and approaches the critical point ($g=1$) linearly.
We consider the initial wave function of the system to be the ground state of $H_{np}(g=0)$ followed by the quench,
\begin{equation}
\label{asha}
g(t)=\frac{1}{\tau_\delta}t;\hspace{0.5cm}0\leq t\leq \tau_\delta =\tau(1+\delta),
\end{equation} 
where  $\tau$ is the total quench time in the absence of 
%without any 
disorder, $\tau_\delta$ is the total quench time in the  disordered case, with $\delta$ being the 
%quantity responsible for the 
disordered parameter. Now, we assume that $\delta$ is a number chosen from a distribution, $p(\delta)$, which is given as 
\begin{eqnarray}\label{chord}
p(\delta)&=&\frac{ \exp\left( \frac{ -\delta^2}{2\sigma^2} \right)}{\sqrt{2\pi}\sigma\text{erf}\left(\frac{3}{\sqrt{2}}\right)};\hspace{0.5cm}|\delta | \leq 3\sigma, \nonumber\\ 
p(\delta)&=&0; \hspace{2.5cm}\text{otherwise,} 
\end{eqnarray}
 where \(\mbox{erf}(\cdot)\) is the Gauss error function.
 The probability distribution is very close to the ubiquitous Gaussian distribution.
The reason for choosing such a distribution, instead of the usual Gaussian, is that $\tau_\delta$ cannot be negative, which  implies that $\delta$ cannot be less than -1.
With the chosen distribution, this is achieved by choosing 
%, which hinders the selection of a gaussian distribution whose tails go to infinity. It is also practically justifiable to use a gaussian distribution with a finite range ($|\delta|<3\sigma$) since probability of $|\delta|>3\sigma$ is very low. Since $\delta < -1$, this implies $3\sigma<1$, therefore 
$\sigma$ in the range %has the range 
$[0,\frac{1}{3}]$. %Note that one can also take distribution with the range of $\delta$ beyond $3\sigma$, e.g., say $4\sigma$, but in that case the range of $\sigma$ to be studied will become narrow. 
Note that $\sigma$ is no more the standard deviation of the distribution $p(\delta)$, but it can still be seen as a measure of dispersion. 

%Note that for the given normal distribution, probability of obtaining a number lesser or greater than $-3\sigma$ or $3\sigma$, respectively, is very close to zero. Therefore, demanding $\tau'$ to be positive, puts restriction on $\sigma$ and thus we  choose $\sigma\leq\frac{1}{3}$.

Solving the equation of motion in Heisenberg picture numerically,
%By solving numerically, 
we observe that for slow quenches $(\omega\tau \gg 1)$, the disorder averaged residual energy follows a power law fall with respect to the quench time $\tau$, i.e., $\overline{E}_r \propto \tau^{\nu}$, 
with $\nu$ remaining close to $-\frac{1}{3}$, whatever be the $\sigma$, 
%of the given distribution 
in the range $\left[0,\frac{1}{3}\right]$. This suggest that the scaling is robust against the 
disorder present in total quench time. 
 We give the scaling exponents, $\nu$, for the scaling relation $\overline{E}_r \propto \tau^{\nu}$, with varying dispersion of the disorder distribution in Table \ref{tab1}. It can be seen that as $\sigma$  increases, there is no change in the scaling exponent $\nu$, correct to the third decimal place, it is still \(-\frac{1}{3}\). %But the change in the exponent is so less that upto its second decimal point $\nu=\frac{-1}{3}$ for $\sigma \in \left[0,\frac{1}{3}\right]$. 

%comment on adiabaticity
\begin{table}[h!]
%\label{maer-paer-jaba-haye}
  \begin{center}
    \caption{\textbf{Response to disorder in quench time on scaling exponent of non-adiabaticity.} We tabulate here the scaling exponents of the disorder averaged residual energy,  $\overline{E}_r$, with the total time of ordered quench, $\tau$,  for different dispersions of the  disorder distribution. The results are obtained by solving the Eq. \eqref{heisen} numerically. Our precision enables us to predict only up to three significant figures. Within this precision, the scalings remain unchanged with change of the disorder strength, as quantified by the dispersion, \(\sigma\). Compare this with the variation in the scalings in the next table. All quantities in the table are dimensionless.}
    \label{tab1}
    \begin{tabular}{|c|c|}
    \hline 
      \textbf{Dispersion} & \textbf{Scaling exponent} \\
      $\sigma$ & $\nu$  \\
      \hline\hline
 0.01 & -0.333 \\[0.8ex]
 \hline
 0.1 & -0.333\\[0.8ex]
 \hline
 0.2 & -0.333\\[0.8ex]
 \hline
 0.3 & -0.333\\[0.8ex]
 \hline
 0.33 & -0.333 \\ [0.8ex] 
 \hline
    \end{tabular}
  \end{center}
\end{table}

\subsection{Scaling exponents within adiabatic perturbation theory and Kibble-Zurek mechanism}
\label{subsec3a}
Adiabatic perturbation theory (APT) is 
a useful method to solve the dynamics of a system  going through a very slow quench \cite{grandi'10}. For a quench given in Eq. \eqref{lata}, with $g_f \ll 1$ and $\omega \tau\gg 1$ (slow quench), it has been shown that the residual energy, within APT, behaves as \cite{plenio'15}
\begin{equation}
\label{manna}
E_r(\tau) \approx \frac{\hslash g^4_f}{16\omega(1-g^2_f)^{\frac{5}{2}}}\tau^{-2}.
\end{equation}
 But APT fails when the system has a vanishing energy gap during the dynamics, even for slow quenches. Therefore, for a quench with $g_f\sim 1$, the scaling given by Eq. \eqref{manna} is no more correct. In such cases, the KZM can be used, and it was shown in \cite{plenio'15}
 that 
 %Thus, the ideas of KZM can be applied to obtain the scaling 
 for $g_f\sim 1$, the  
 %and it is shown that on the application of KZM the 
 scaling exponents within KZM come out to be $-\frac{1}{3}$,
just as obtained numerically from the equation of motion in the Heisenberg picture directly. 
%which was obtained numerically, too.

In this subsection, our aim is to verify whether for disordered systems, scalings predicted by KZM matches with that obtained directly from the Heisenberg equation of motion (as already presented in Table \ref{tab1}). 
Consider, therefore, the quench given in Eq. \eqref{asha}.
 The basic idea of KZM is to split a dynamics into adiabatic and impulsive  regimes based on a comparison of two time scales \cite{dziar'10,zurek'14,damski'08,plenio'17}. The first of these time scales is the ``relaxation time'' of the system, which is inversely proportional to the energy gap ($\eta$) between the ground state and the first allowed excited state \cite{zurek'05}. Precisely, it is
 %, i.e. 
 $\hslash \eta^{-1}$, where $\eta=2\hslash \omega \sqrt{1-g^2}$. The relaxation time of the system is the time required by the system to respond to a given quench. And the second one, which is known as the ``transition time'' \cite{damski'07}, is the time scale in which there occurs changes in the Hamiltonian, and is characterized by $|\eta/ \dot{\eta}|$. If the relaxation time is lower than the transition time, the dynamics of the system is in the adiabatic regime, but otherwise, 
 %if the relaxation time is greater than the latter time scale then 
 the dynamics is in the impulsive regime and the system stops to react to the quench. 
 
  Let the value of the quench parameter which splits the adiabatic and impulsive regimes be given by $g=\hat{g}$, which can be obtained using
\begin{equation} \label{kishore}
\left|\frac{\eta(\hat{g})}{\dot{\eta}(\hat{g})}\right|=\frac{\hslash}{\eta(\hat{g})},
\end{equation}
whence we  get
\begin{equation}
\label{Akhtaribai}
\frac{(1-\hat{g}^2)^{3/2}}{\hat{g}}=\frac{1}{2\omega \tau_\delta}.
\end{equation}
%We remember that $\tau_\delta$ depends on a  disorder parameter $\delta$, which is chosen from a distribution given in Eq. \eqref{chord} of mean zero and a dispersion $\sigma$. Also, we assumed that the $\sigma$ is so small such that $\tau_\delta$ cannot be negative. This implies $\sigma$ is so chosen that  $\delta>-1$ is violated only for an insignificant amount of time (see Eq. \eqref{asha}).
Note that for $\sigma \in [0,0.33]$ (range taken in Table \ref{tab1}),  for the distribution given in Eq. \eqref{chord}, $\delta$ is never closer enough to -1, so that $(1+ \delta)^{-1}$ blows up.
Therefore, for a slow quench ($\omega\tau \gg 1$) and a smaller enough dispersion $\sigma$, $\omega\tau_\delta \gg 1$, too. 
%\textcolor{red}{[But, \(\delta\) can be close to -1, so that \(\tau_\delta\) can be close to 0. Then $\omega\tau_\delta \gg 1$ won't hold anymore!!]} 
This implies that the R.H.S. of Eq. \eqref{kishore} is close to zero, and 
L.H.S. is close to zero when $\hat{g}\approx 1$, whereby we can write \(\hat{g} = 1- \hat{\epsilon}\), for a small positive \(\hat{\epsilon}\). Ignoring higher than linear order in \(\hat{\epsilon}\), we 
obtain \(\hat{\epsilon} \approx 1/(2/3 + 2^{5/3}(\omega \tau_\delta)^{2/3})\), and because \(\omega \tau_\delta \gg 1\), we get
%
%therefore it can be re-expressed as
%&1&- \hat{g}\simeq \frac{1}{2^{5/2}(\omega \tau_\delta)^{2/3}}, \nonumber \\
%\implies &
\begin{equation}
\label{Akhilbandhu}
\hat{g} \approx 1-\frac{1}{2^{5/2}(\omega \tau_\delta)^{2/3}}.
\end{equation}

%Let us now denote the residual energy at the end of the disordered quench given in Eq. \eqref{asha}, for a particular realization  \(\delta = \delta\) of the disordered parameter, by $E_r(\tau,\delta)$. 

Since $\hat{g} \lesssim  1$, and the system freezes at $g=\hat{g}$, therefore, $E_r(\tau_\delta) \approx E_r(\hat{g})$, where \(E_r(\hat{g})\) is the residual energy when \(g=\hat{g}\). 
Now, it can be shown using APT that for slow quenches $(\omega\tau_\delta \gg 1)$, the residual energy, $E_r(\hat{g})$, is given by
\begin{equation}
E_r(\hat{g}) \approx \frac{\hslash \hat{g}^2}{16\omega(1-\hat{g}^2)^{5/2}}\tau^{-2}_\delta,
\end{equation}
which, using relations 
(\ref{Akhtaribai}) and (\ref{Akhilbandhu}), can be expressed as 
\begin{equation}
E_r(\hat{g}) \approx  \frac{\hslash \omega}{4 \sqrt[3]{2} (\omega \tau)^{\frac{1}{3}}}  \left[ (1+\delta)^{-\frac{1}{3}} - \frac{(1+\delta)^{-1}}{12 \sqrt{2}(\omega \tau)^{\frac{2}{3}}}\right].
\end{equation}
%
%
%\nonumber \\
%&=&\frac{\hslash\omega}{8}\left[2^{5/2}(\omega\tau)^{-1/3}(1+\delta)^{-1/3}-(\omega\tau)^{-1}(1+\delta)^{-1}\right]. \nonumber \\
%\end{eqnarray}
The disorder averaged residual energy, at the end of the quench, is given as $\overline{E}_r=\int E_r(\hat{g}) p(\delta) d\delta $. Now, the integrals of $(1+\delta)^{-1/3}p(\delta)$ and $(1+\delta)^{-1}p(\delta)$ are of the order of 1, for $\sigma<0.33$.
% \sout{       \textcolor{red}{Now, the integrals of $(1+\delta)^{-1/3}e^{-\delta^2/{2\sigma^2}}$ and $(1+\delta)^{-1}e^{-\delta^2/{2\sigma^2}}$, are of the order of 1 for $\sigma<0.2$. Note that higher values of $\sigma$ are forbidden as already mentioned before, and moreover we find here that there exist  singularities in the integrands when $\delta=-1$ for $\sigma \geq 0.2$.} \textcolor{red}{[Is this true? What does it mean by ``singularity in the integrand''? The singularity in the integrand is there for all \(\sigma\)! No? Maybe due to integration that singularity becomes finite for \(\sigma \geq 0.2\)?]}     }
%and the results of the integrals are complex numbers.
 Therefore the disorder averaged residual energy for low values of $\sigma$ is
\begin{equation}
\overline{E}_r \approx \frac{\hslash\omega}{4\sqrt[3]{2}}(\omega\tau)^{-1/3}.
\end{equation}
 Note that the second term in the expression of $\overline{E}_r$ is dropped because of the slow quench, i.e., $\omega\tau \gg 1$.
Therefore the scaling exponent obtained using APT and KZM is $\nu=-\frac{1}{3}$, which matches with the numerically obtained data by directly solving the equation of motion in the Heisenberg picture, as presented in Table \ref{tab1}.  
  Thus, both the analytical but approximate calculation using APT and KZM as well as the numerical but exact calculation provide the same scaling for the averaged residual energy in the disordered dynamics in which quench stops at the critical point of the ordered quantum Rabi model, and matches with the scaling obtained in the ordered case. The disorder considered is of quenched type and is inserted in the total time of quench. We will now investigate the outcome of insertion of disorder in the final value of the quench parameter.

\begin{figure}[t!]
\includegraphics[width = 0.3\textwidth, angle=-90]{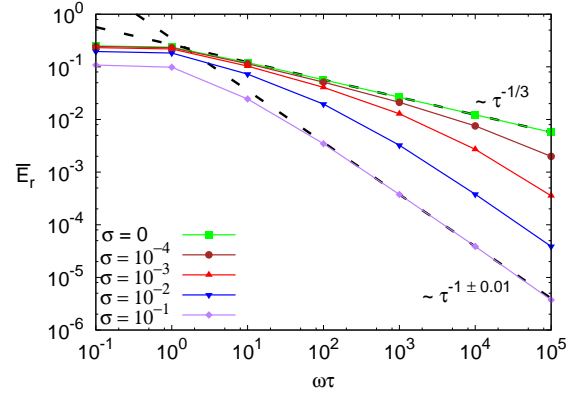} 
\caption{\textbf{Analyzing adiabaticity in disordered quenches in the quantum Rabi model.} The disorder averaged residual energy, $\overline{E}_r$, is plotted against $\omega\tau$ of a disordered quench (given in Eq. \eqref{rafi}). For slow quenches $(\omega\tau \gg 1)$ and small standard deviation ($\sigma$) of the disorder, $\overline{E}_r$ maintains a power-law scaling with the total time of quench, $\tau$, of the corresponding ordered quench. As  $\sigma$ shifts from 0 to 0.1, 
%a transition in 
the scaling exponent shifts 
%is observed, i.e., 
from $-\frac{1}{3}$ to $-1\pm0.01$.  For the trend of the scalings with varying $\sigma$, see also Table \ref{tab2}. The degrees of adiabaticity of the quenches increase with the increase in strength of the disorder, as quantified by the standard deviation, $\sigma$, of the disorder. $\overline{E}_r$ is plotted in the units of $\hslash\omega$ and $\omega\tau$ is dimensionless.
}
\label{fig2}
\end{figure} 

\section{Dynamics with disorder in quench parameter}
%Quench ending at Disordered Hamiltonian}
\label{sec4}
In this section, we consider the quench given in Eq. \eqref{lata} with the disorder in the final Hamiltonian, i.e.,  the final value of the quench parameter, $g$.
 The disorder is chosen in such a way that it offers resistance to the quench to end at the critical point (while ``approaching'' from the left, i.e. over lower values), and thus the disordered final value of quench parameter, $g$, is given by
\begin{equation}
g_f(\delta)=(1-|\delta|), \nonumber
\end{equation}
where, in this case, the disorder parameter, $\delta$, is chosen from a quenched Gaussian distribution of mean zero and standard deviation, $\sigma$. Therefore the quench is given by
\begin{equation}\label{rafi}
g(t)=\frac{g_f(\delta)}{\tau}t;\hspace{1cm}0\leq t\leq \tau.
\end{equation}

 Also, $\sigma$ is taken in the range $[0,~0.1]$, resulting in the system to have $g(t)>0$ satisfied with very high probability. We numerically obtain the disorder averaged residual energy, $\overline{E}_r$, and plot it in the units of $\hslash\omega$, with respect to  $\omega\tau$, in Fig. \ref{fig2}. 
  It is known that for the case with no disorder, i.e., $\sigma=0$, $\overline{E}_r\propto \tau^{-\frac{1}{3}}$, for higher values of $\omega \tau$. For the case with the disorder, it is observed that for $\sigma = 0.1$, again a power law appears such that $E_r \propto \tau^\nu$, where $\nu=-1\pm0.01$ (see Table \ref{tab2}). Both the cases are depicted by a linear-fit in the Fig. \ref{fig2}.  Therefore, as we scan from  $\sigma=0$ to $\sigma=0.1$, there is a transition in the $E_r$ versus $\omega \tau$ behavior. In order to demonstrate this transition, we estimate the scaling exponents, $\nu$, as a function of $\sigma \in [0,~0.1]$, for $\omega \tau \in [10^3,~10^4]$ and  $\omega \tau \in [10^4,~10^5]$. 
 The scaling exponents for various $\sigma$ are presented in Table \ref{tab2}.
 One can see that as  $\sigma$ increases, $\nu$ decreases, indicating the increase in the adiabaticity of the quench. This feature is expected, because if the quench stops farther from (and before reaching) the critical point, then adiabaticity should improve.

%  For larger values of $\omega\tau$ and for higher enough $\sigma$, i.e., $\sigma \approx 0.1$,  $\overline{E}_r$ starts to follow a power law with $\tau$, i.e., $\overline{E}_r \propto \tau^\nu$, where $\nu$ is the scaling exponent. For the case when $\sigma \approx 0.1$, and $\nu=-0.99$, correct upto second decimal point,  for $\omega\tau$ to be in the range $[10^3,~10^4]$. Note that the scaling exponents are not same for the values between $\sigma=0$ to 0.1, except at the extreme ends, this indicates that the power-law behavior starts to appear only for larger enough $\sigma$ of the disorder.

 \begin{table}[t!]
  \begin{center}
    \caption{\textbf{Response to disorder in quench parameter on scaling exponent of non-adiabaticity.} 
    The considerations are the same as in Table \ref{tab1}, except that the disorder is inserted in the quench parameter \(g\), and the quench is 
    %. 
    %Scaling exponents with the standard deviation, for the disordered quench 
    given by Eq. \eqref{rafi}. All quantities in the table are dimensionless.}
    \label{tab2}
    \begin{tabular}{| m{4em} | m{2cm}| m{2cm} |}
    \hline 
      \textbf{Standard deviation} $\sigma$ & \textbf{Scaling exponent} for $\omega\tau \in [10^3,~10^4]$ & \textbf{Scaling exponent} for $\omega\tau \in [10^4,~10^5]$\\
%      $\sigma$ & $\nu$ &  \\
      \hline\hline 
     $0$ & -0.33 & -0.33\\ [0.8ex]
 \hline
      $10^{-4}$ & -0.44 & -0.57\\ [0.8ex]
 \hline
 $10^{-3}$ & -0.67 & -0.87\\ [0.8ex]
 \hline
 $10^{-2}$ & -0.93 & -0.98\\ [0.8ex]
 \hline
 0.1 & -0.99 & -1.01\\ [0.8ex]
 \hline
% 0.2 & -0.988411\\ [0.8ex] 
% \hline
    \end{tabular}
  \end{center}
\end{table}

\begin{figure}
\begin{subfigure}{.22\textwidth}
  \centering

  \includegraphics[width=.8\linewidth, angle=-90]{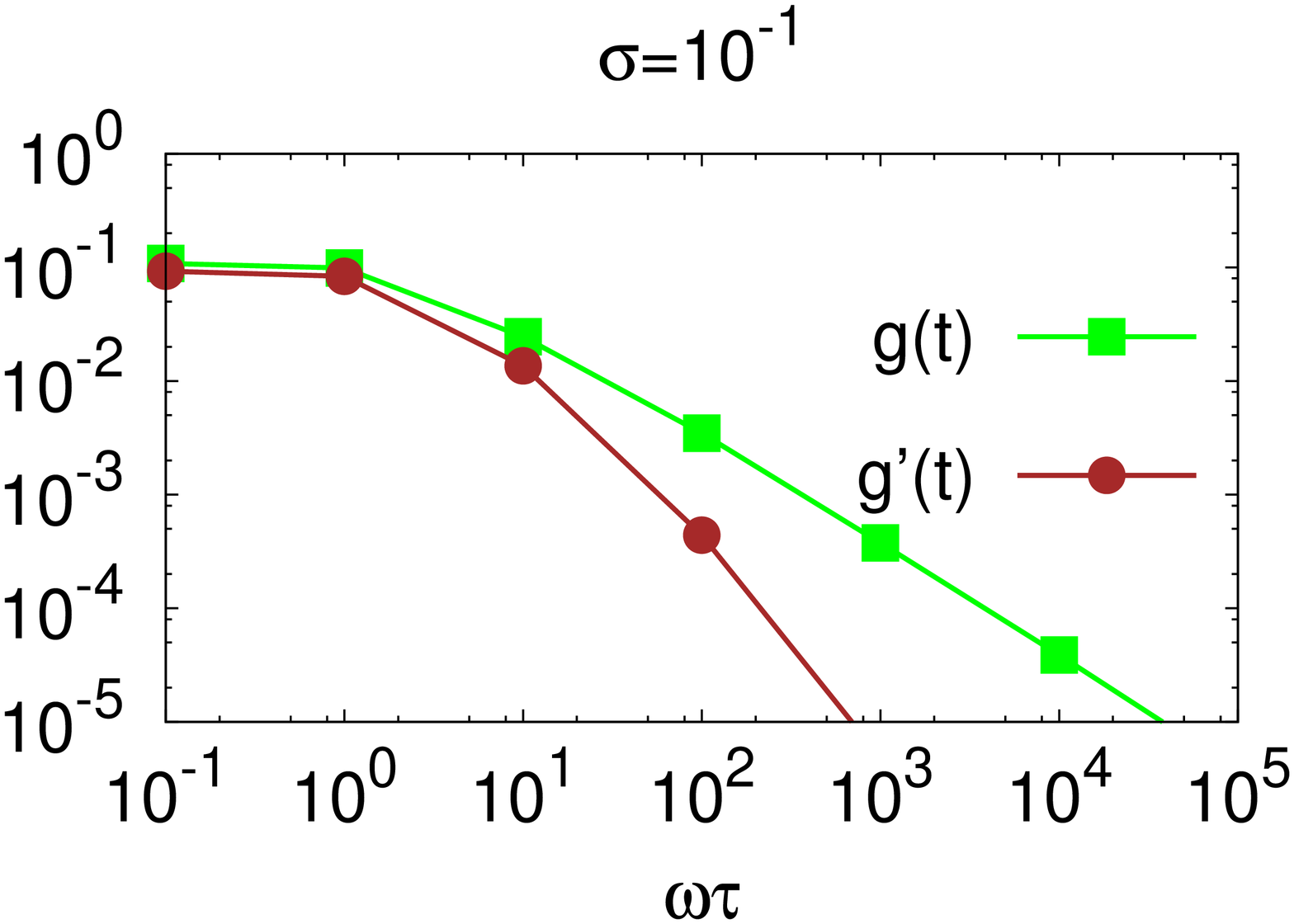}  
  \label{fig:sub-first}
\end{subfigure}\hspace{0.5cm}
\begin{subfigure}{.22\textwidth}
  \centering
  \includegraphics[width=.8\linewidth, angle=-90]{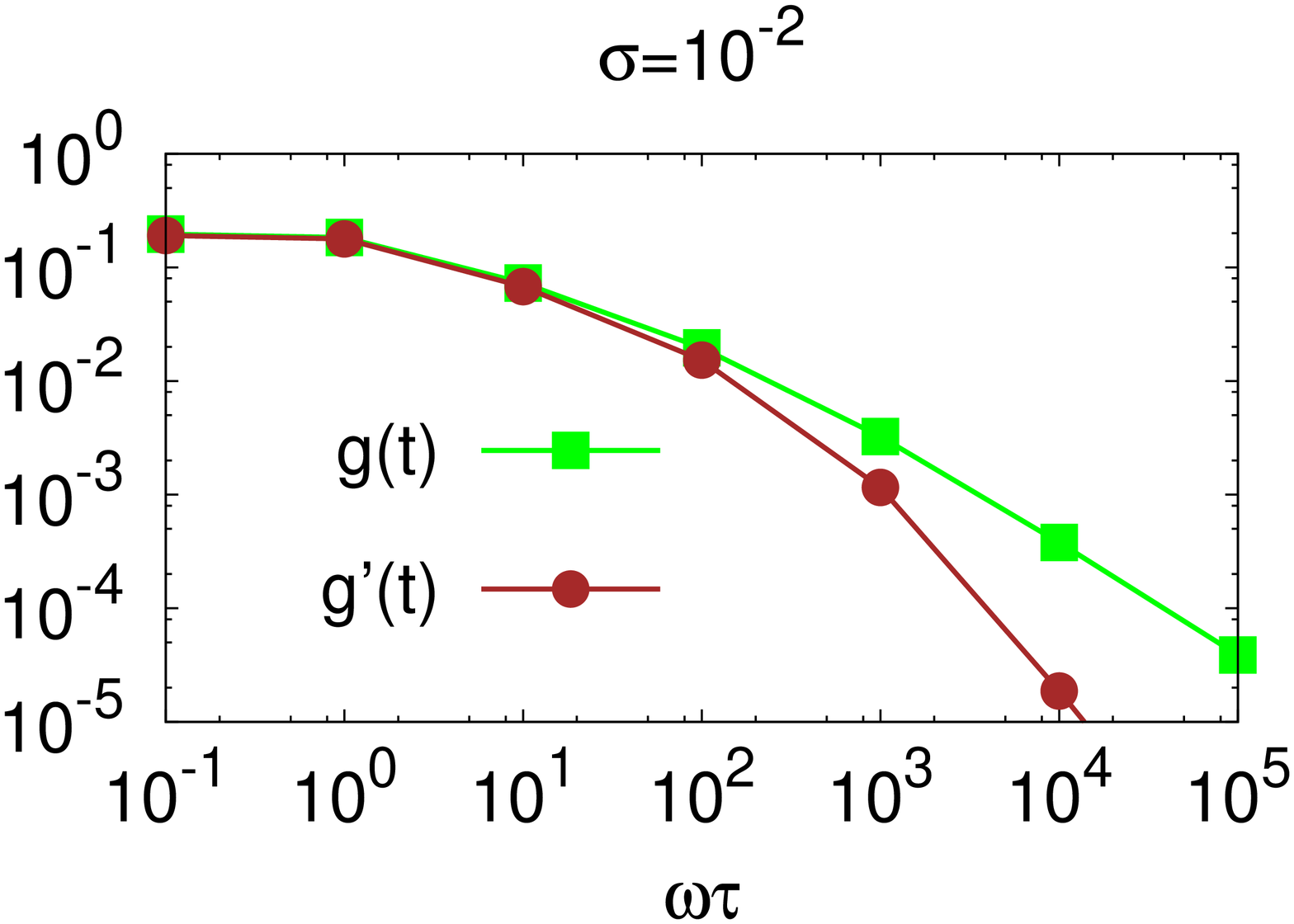}  
  \label{fig:sub-second}
\end{subfigure}

\begin{subfigure}{.22\textwidth}
  \centering
 
  \includegraphics[width=.8\linewidth, angle=-90]{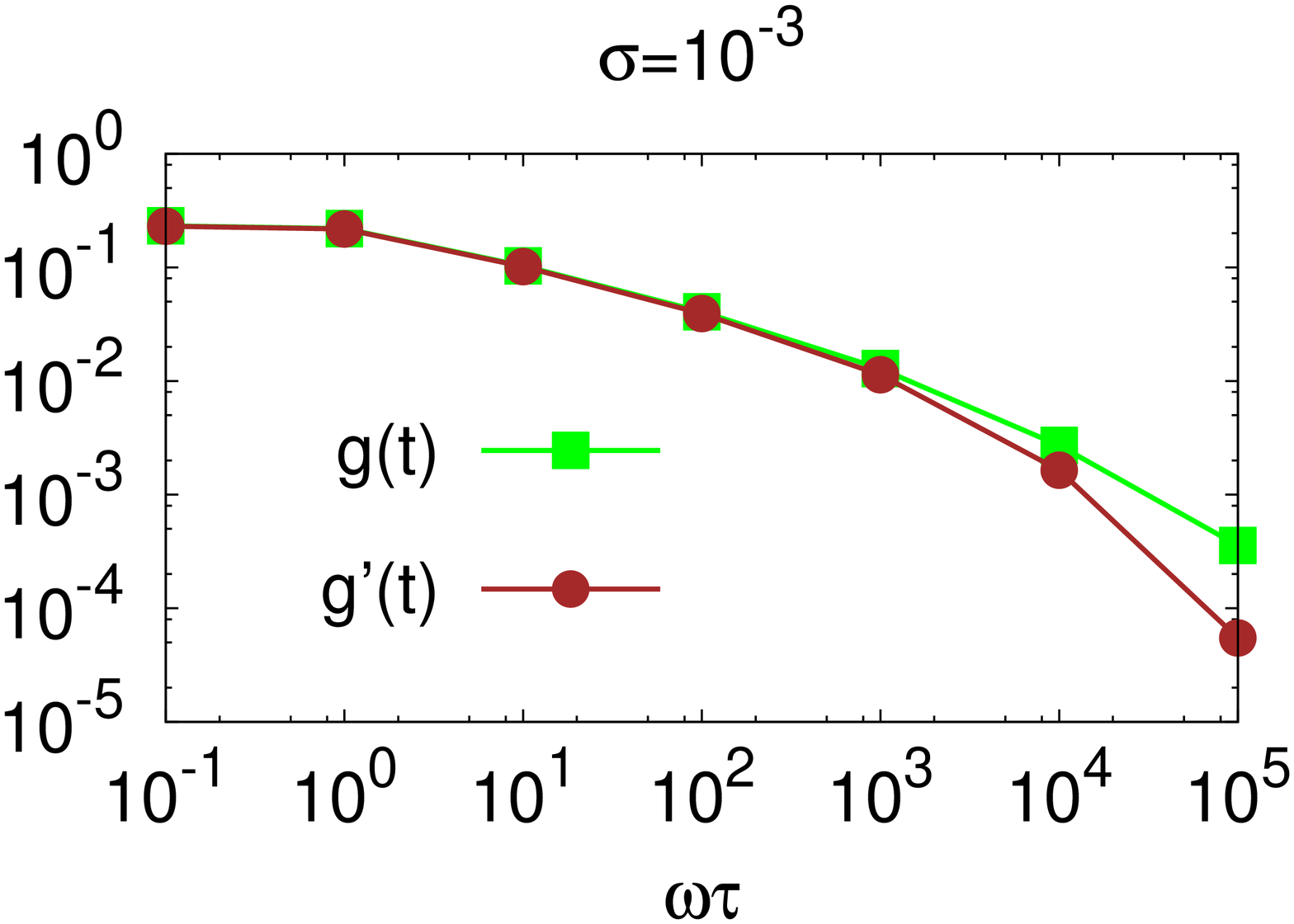}  
  \label{fig:sub-third}
\end{subfigure}\hspace{0.5cm}
\begin{subfigure}{.22\textwidth}
  \centering
  \includegraphics[width=.8\linewidth, angle=-90]{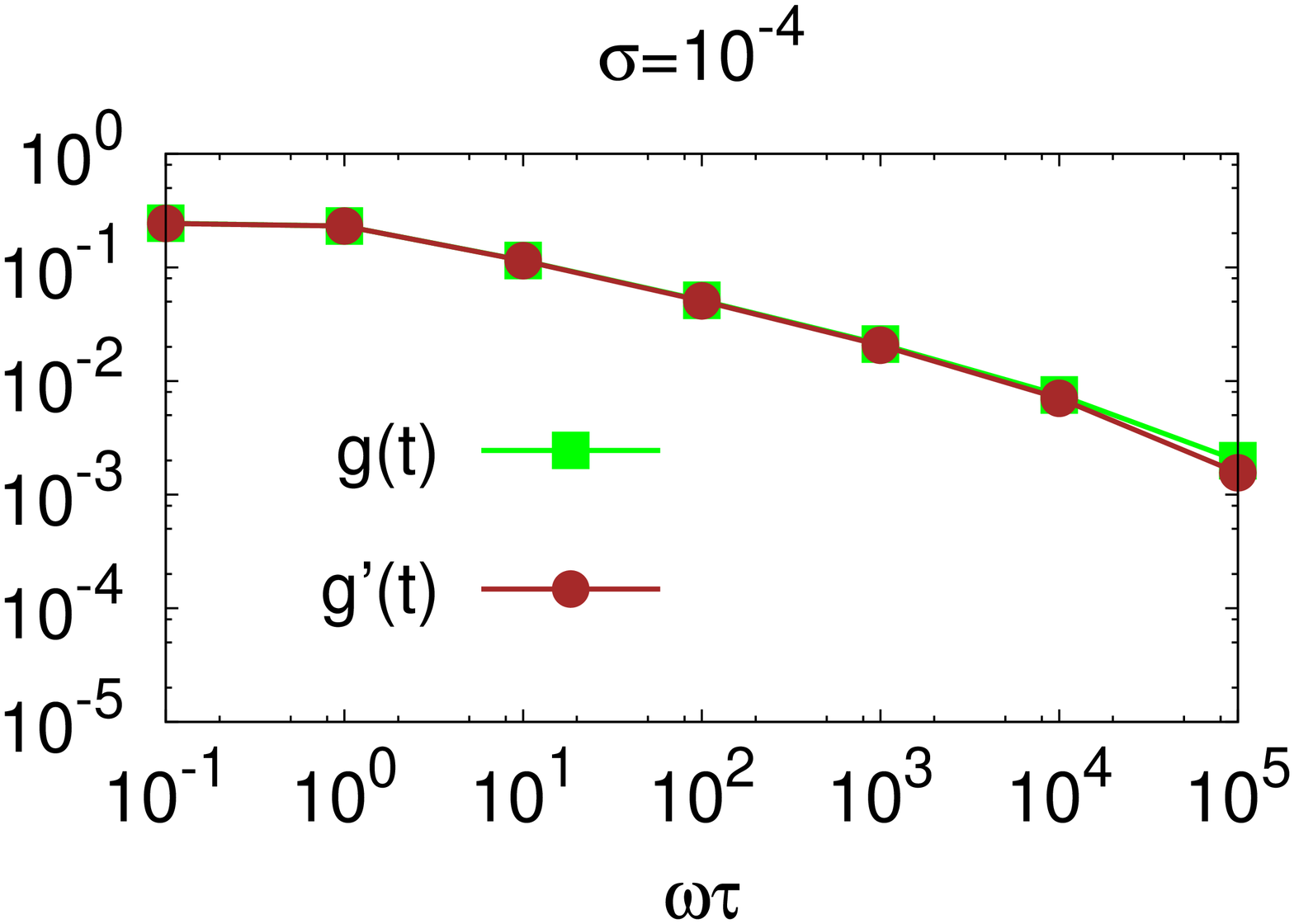}  
  \label{fig:sub-fourth}
\end{subfigure}
\caption{\textbf{Comparing disorder-averaged non-adiabaticity with non-adiabaticity for disorder-averaged quench parameter.} Each panel is for a different standard deviation, \(\sigma\), as mentioned above the panel. For a given \(\sigma\), the disorder averaged residual energy, $\overline{E}_r$, is plotted  on the vertical axis, against $\omega\tau$ on the horizontal axis, for the disordered quench, $g(t)$, %and $g'(t)$ 
given by Eq. \eqref{rafi}. This is compared with the residual energy (plotted on the same vertical axis) for the disorder-averaged quench, $g'(t)$, given by \eqref{sonu}.
%
%
%respectively, for various $\sigma$.  
$g'(t)$ offers a higher degree of adiabaticity than $g(t)$. The difference in the degree of non-adiabaticities becomes smaller with decrease in $\sigma$. $\overline{E}_r$ is in the units of $\hslash\omega$ and $\omega\tau$ is dimensionless.}
\label{fig3}
\end{figure}

In order to analyze the effect of disorder further, we define a quantity which is the disorder average of $g_f(\delta)$, i.e., \begin{equation}
\overline{g}_f(\sigma)=\int_{-\infty}^\infty g_f(\delta). \frac{1}{\sigma\sqrt{2\pi}}e^{-{\delta^2}/{2\sigma^2}}d\delta = 1-\sqrt{\frac{2}{\pi}}\sigma,
\end{equation} and consider a quench, 
\begin{equation}\label{sonu}
g'(t)=\overline{g}_f(\sigma)\frac{t}{\tau}; \hspace{1cm} t\leq \tau, 
\end{equation}
in the Hamiltonian $H_{np}(g'(t)).$
Note that now the final Hamiltonian of the quench is $H_{np}(\overline{g}_f(\sigma))$. Since $\overline{g}_f(\sigma)<1$ for non-zero $\sigma$, this is also, therefore, a case of quench ending before reaching the critical point. The standard deviation $\sigma$ plays the role of distance between the end point of the quench parameter, $\overline{g}_f(\sigma)$, and the critical point $g'(t)=1$. 
Let us denote the scaling exponent in this case to be $\nu'$ for $\omega\tau \in [10^3,~10^4]$.
It is observed that the scaling exponent $\nu'<\nu$ for non-zero standard deviation $\sigma$ of the disorder. We present the numbers in the Table \ref{tab3} and Fig. \ref{fig3}. This indicates that the adiabaticity is higher when considering a disorder averaged quench $g'(t)$,  than the disorder averaged adiabaticity.
%disordered case considered in Eq. \eqref{rafi}.  
The difference between the scale exponents of the  residual energies for disorder averaged quench and the averaged residual energy for the disordered quench decreases with decrease in the standard deviation, $\sigma$, of the disorder.

\begin{table}[h!]
  \begin{center}
    \caption{\textbf{Scaling of disorder-averaged non-adiabaticity versus the same of non-adiabaticity in disordered quench.} We present here a comparison, for different standard deviations, between the scaling exponents \(\nu\) and \(\nu{'}\),
    respectively 
    %for with the standard deviation, 
    for the quench, $g(t)$, given by Eq. \eqref{rafi}, and $g'(t)$, given by Eq. \eqref{sonu}. All quantities in the table are dimensionless.}
    \label{tab3}
    
      \begin{tabular}
{|c|c|c|}      
\hline
      $\sigma$ & $\nu$ & $\nu'$ \\
      \hline\hline
       $0$ & -0.33 & -0.33\\[0.8ex] 
 \hline
      $10^{-4}$ & -0.44 & -0.46\\[0.8ex] 
 \hline
 $10^{-3}$ & -0.67 & -0.84\\[0.8ex]
 \hline
 $10^{-2}$ & -0.93 & -1.79\\[0.8ex]
 \hline
 $0.1$ & -0.99  & -1.98\\[0.8ex]
 \hline
% 0.2 & -0.988411 & -1.99378\\ [0.8ex] 
% \hline
    \end{tabular}
  \end{center}
\end{table}

\section{Conclusion}
\label{sec5}

Any system driven close to a phase transition unleashes distinct features compared to when its dynamics is considered far away from the same. In particular, a dynamics far away from its critical points of phase transition can be adiabatic for a sufficiently slow quench, whereas this is not possible for quenches near the critical points.
Non-adiabaticity for slow quenches in quantum Rabi model has been analyzed in earlier in the literature.  The quantum Rabi model, a simplified form of the Dicke model, possesses  normal and superradiant phases in the limit when the ratio of atomic transition to cavity field frequencies diverges. We have studied the non-adiabaticity of slow quenches in presence of disorder in the quench of the system governed by the Rabi Hamiltonian. We considered a disordered version of the quench in which  system residing in the ground state of an initial Hamiltonian of the normal phase is quenched to the final Hamiltonian corresponding to the critical point. The disorder was assumed to exist either  in the total time of the quench or in the quench parameter itself. 
%rate of the quench or in the final  Hamiltonian of the system which does not correspond to critical point. 
We numerically simulated the corresponding Schr{\"o}dinger equations, and obtained that the non-adiabatic effects are unaffected - in terms of the scaling exponents - by the presence of disorder in the total time of the quench. 
We also independently checked the scaling exponents by using adiabatic perturbation theory and the Kibble-Zurek mechanism.
 For the case when the disorder is in quench parameter, a monotonic increase in the adiabaticity is reported with the strength of the disorder. We subsequently also considered a quench where the final Hamiltonian is chosen as average over the disordered final Hamiltonians with the disorder being in the quench parameter. We showed that this quench is more adiabatic than the average of the quenches with the disorder in final Hamiltonian quench parameter.

\end{document}